\documentclass[final,5p,times,twocolumn]{elsarticle}
\usepackage{graphicx}
\usepackage{subfigure}
\usepackage{epstopdf}
\usepackage{color}
\usepackage{multirow}
\usepackage{textcomp,gensymb}
\usepackage{lineno}
\usepackage{xcolor}
\usepackage{epsfig}
\usepackage{amsmath}
\usepackage{amssymb}
\usepackage{booktabs}
\usepackage{siunitx}
\usepackage{breqn}
\biboptions{square,comma,sort&compress}
\journal{}
\begin{document}
\begin{frontmatter}
\title{Solute Segregation in a Moving Grain Boundary: A Novel Phase-Field Approach
}
\author[a,b,d]{Sandip Guin}
\author[c]{Miral Verma}
\author[a]{Soumya Bandyopadhyay}
\author[b,d]{Yu-Chieh Lo*\corref{cor}}
\ead{yclo@nycu.edu.tw}
\author[a]{Rajdip Mukherjee*\corref{cor}}
\ead{rajdipm@iitk.ac.in}
\address[a]{Department of Materials Science and Engineering, Indian Institute of
Technology, Kanpur, Kanpur-208016, UP, India}
\address[b]{International College of Semiconductor Technology, National Yang Ming Chiao Tung University, Hsinchu 300, Taiwan} 
\address[c]{Department of Materials Engineering, KU Leuven, Kasteelpark Arenberg 44, Leuven 3001, Belgium}
\address[d]{Department of Materials Science and Engineering, National Yang Ming Chiao Tung University, Hsinchu 300, Taiwan}

\begin{abstract}
\textcolor{black}{We present a novel phase-field approach for investigating solute segregation in a moving grain boundary. In our model, the correct
choice of various parameters can control the solute-grain boundary interaction potential, resulting in various segregation profiles that agree with Cahn’s solute drag theory. Furthermore, we explore how different segregation profiles evolve at varying GB velocities owing to the inequality of the atomic flux of solute between the front and back faces of the moving grain boundary. We highlight velocity variations among segregation profiles in low and high-velocity regimes. This model reveals how grain boundary segregation affects grain growth, providing insights for future alloy design.}

%subject classification numbers as needed.
\end{abstract}
\begin{keyword}
Grain boundary segregation \sep solute drag \sep phase-field model  

\end{keyword}
\end{frontmatter}

%\linenumbers

%\section{Introduction}
%\label{sec:sample1}

\section{Introduction}
Grain boundary (GB) segregation, observed in poly-crystalline materials, involves solute atom accumulation in GBs, leading to the stabilization of grain structure~\cite{Lejcek201783, Raabe2014253, MILLETT20072329, KIRCHHEIM2002413}. 
%Segregation in the GB leads to a reduction in grain boundary energy and solute drag effect, manifesting slower grain growth~\cite{HILLERT1976731}\cite{CAHN1962789}\cite{CANTWELL20141}. Studies report that researchers have utilized GB segregation to prevent grain growth in nano-crystalline materials~\cite{D0NR07180C}\cite{LI2015110}\cite{MILLETT20072329}. Recent studies suggest solute segregation in boundary leads to a phase transformation at Gbs~\cite{RAABE20136132}. In some cases, it causes grain boundary spinodal decomposition~\cite{KWIATKOWSKIDASILVA2019109}.
%The most challenging aspect is to quantify the amount and nature of grain boundary segregation. There is a complex coupling between grain boundary solute segregation with different thermodynamics parameters like the chemical potential of each element, atomic size difference, grain boundary energy, temperature, etc. In the last few decades, researchers have developed many theories to quantify grain boundary segregation~\cite{MPSeah1980}\cite{PavelLe}. 
 For example, solute segregation stabilizes the ultrafine grain structure in cold-sprayed 6061 aluminium powder particles~\cite{ROKNI2014482}. Segregation reduces grain growth through two mechanisms: (i) solute segregation at GBs lowers GB energy~\cite{KIRCHHEIM2002413,AKSYONOV2017266,HU2020109271}, and (ii) solute drag force by GB solute impedes GB migration~\cite{CAHN1962789,TODACARABALLO201395}.  Moreover, this solute drag phenomenon in migrating grain boundaries depends on various factors such as grain boundary mobility, solute diffusivity, misfit strain, etc.~\cite {PhysRevLett.127.175503,2021}. 
%Even a minute change in the segregation controlling parameters can significantly alter the effect of solute drag.
%Furthermore, the complex interplay between grain boundary solute segregation and different  thermodynamic parameters such as the chemical potential of each element, atomic size difference, 
%grain boundary energy, temperature, etc. makes it a more challenging to quantify the amount  and nature of grain boundary segregation~\cite{LEGALL19994365}.

Over the last decade, researchers have conducted experimental investigations to quantify GB segregation. \emph{Lejcek et al.} explored temperature-dependent segregation of Sn and Sb in BCC iron GBs, which can be elucidated by both segregation enthalpy and entropy considerations~\cite{LEJ}. \emph{Xie et al.} found that Sn segregation in Zircaloy-4 alloys is influenced by the crystallographic orientation~\cite{XIE2016225}. Using first principle approach \emph{Umashankar et al.} studied Y, Zr and Nb segregation in bcc TI-Mo alloy~\cite{UMASHANKAR2023112393}.  Additionally, \emph{Zhang et al.} delved into the mechanisms of phosphorus segregation in steel through the use of Auger electron spectroscopy (AES)~\cite{ZHANG2015171}.

Over recent decades, researchers have 
proposed numerous theories to quantify 
grain boundary (GB) 
segregation~\cite{PavelLe, LUCKE19711087}. 
%The prevailing Langmuir-McLean isotherm 
%theory is commonly employed 
%to measure GB segregation extent. 
In a 
comprehensive monograph, \emph{J. W. Cahn} 
introduced the solute drag theory, 
elucidating equilibrium GB segregation in 
both stationary and moving GBs based on 
solute-GB interaction 
energy~\cite{CAHN1962789}. Cahn's article 
establishes the relationship between GB 
velocity and drag force. 
%Subsequently, numerous models were 
%developed to study the solute drag effect 
%in migrating grain boundaries.
To explore the influence of segregation on 
grain growth kinetics and size 
distribution, \emph{Fan et al.} employed a phase-field model to simulate grain growth and 
capture the drag effect~\cite{dfan}. 
\emph{Heo et al.} investigated the impact of misfit strain 
on solute-GB interaction using a phase-field model~\cite{HEO20117800}.

In existing models, grain boundary solute 
segregation is typically assumed to be 
mono-layered~\cite{dfan,HEO20117800}. However, recent experimental 
findings indicate a more intricate 
phenomenon, with multilayer segregation 
observed in metallic 
alloys~\cite{GUPTA20073131, CANTWELL20141, 
Nano}. Moreover, most phase-field 
simulations regarding the GB segregation 
assume a symmetrical decrease from the 
peak solute concentration at the middle of 
the GB region~\cite{HEO20117800}. This assumption limits 
their applicability when addressing 
multilayer GB segregation. A broad spectrum of metallic alloys has been experimentally observed to exhibit multilayer segregation~\cite{GUPTA20073131,CANTWELL20141}. \emph{Alkayyali et al.} propose a solute drag model based on regular solutions approach, which incorporates the multilayer grain boundary segregation~\cite{PRL_main}. Their results show that GB-Solute interaction plays significant role in determining solute drag force and multilayer segregation exerts larger drag force than monolayer segregation. \emph{Cahn et al.} showed the solute drag force depends on GB velocity~\cite{CAHN1962789}. Thus, to 
accurately capture the complexities of 
solute drag on moving grain boundaries at different velocity regime and 
account for multilayer GB segregation, 
advanced modeling approaches are crucial.

Thus, in this article, we introduce a 
novel phase-field model to study various 
solute segregation patterns at grain 
boundaries that mimics that multilayer 
segregation.
Phase-field modeling is an elegant 
computational tool for microstructure 
modelling without explicit tracking of 
interfaces~\cite{MATTOSFERREIRA2023112368}. Continuous representation of 
field variables and a diffuse interface 
region enable accurate simulation of 
complex interactions between GBs, surfaces 
and interphase interfaces.  For example, 
\emph{Mukherjee et al.} employed a phase-field 
model to study different pathways of 
microstructure evolution during sintering 
of nanoporous aggregates for different 
combinations of surface, GB diffusivity 
and GB mobility~\cite{rajdipm}.  \emph{Mukherjee et al.} also uses phase field modeling to study GB grooving, revealing anisotropic effects and dynamics~\cite{MUKHERJEE_GB_grove}. \emph{Verma et 
al.} investigated grain growth stagnation 
in solid-state thin films using a phase-
field approach, where the surface grooves 
effectively pin the migrating GBs
~\cite{miral_grain}. In a follow up \emph{Verma et al.} also 
presented a computational analysis of the 
time independent behavior of a thermal 
groove in a migrating GB using a phase-
field model~\cite{miral_grove}. \emph{Suhane et al.} studied the effect of alloying elements in austenite grain growth using phase-field simulations~\cite{SUHANE2023112300}.

\section{Phase-field model}
In this work, we extend the phase-field 
model proposed by \emph{Heo et al.} modifying the parameter controlling the grain boundary and solute interaction potential~\cite{HEO20117800}. 
Our phase-field approach employs a conserved 
order parameter $c$ for solute 
concentration and a set of non-conserved 
parameters $\eta_i; (i = 1 \dots N)$ to 
represent individual grains in a binary 
alloy~\cite{HEO20117800}. The total free 
energy ($F$) of the system is defined as:
    \begin{equation}
          \begin{aligned}
          F_{total} = \int_{V}\Big[f_{chem} +c\cdot E  + {\omega} \cdot g({\eta_1},{\eta_2},..,{\eta_g}) + \frac{\kappa_c}{2}|\nabla c|^2 + \\  \sum_{g}\frac{\kappa_{\eta}}{2}|\nabla {\eta_i}|^2 \Big]dV. 
          \label{eq:seg_2}
          \end{aligned}
    \end{equation}

Here, $f_{chem}$ represents the chemical 
free energy density, %of the solid solution's incoherent state, 
E represents the chemical interaction 
between the solute atoms and the grain 
boundary, and ${g({\eta_1},{\eta_2},..,
{\eta_g})}$ characterizes the local free 
energy density of the grain structure. The 
parameter ${\omega}$ is a constant, and 
${\kappa_c}$ and ${\kappa_{\eta}}$ are the
gradient energy coefficients of 
composition ${c}$ and the grain order 
parameters ${\eta_i}$, respectively. 
We employ a regular solution-based model 
to define the ${f_{chem}}$ in a binary 
system which is given as,

\begin{equation}
    \begin{aligned}
    f_{chem} = \mu^0c + {{\mu^0}_h}(1-c)+RT[clnc+(1-c)ln(1-c)] \\ +\Omega c(1-c).
    \end{aligned}
    \label{eq:finc}
    \end{equation}

Here, ${\mu^0}$ and ${{\mu^0}_h}$ 
represent the standard chemical potential 
for solute and host atoms, respectively. 
${R}$ denotes the gas constant, ${T}$ 
represents the temperature, and ${\Omega}$ 
signifies the regular solution interaction 
parameter. For simplicity, we take 
$\Omega$ to be 0. 

The 
expression for the grain boundary free 
energy density is given below:

\begin{equation}
          \begin{aligned}
          g({\eta_1},{\eta_2},....,{\eta_g}) = &0.25+\sum_{i} \Big (-\frac{{\eta_i}^2}{2}+\frac{{\eta_i}^4}{4} \Big )+ \\ & \gamma \sum_{i}\sum_{j>i}{{\eta_i}^2}{{\eta_{j}}^2},
          \end{aligned}
          \label{eq:fan_chen}
\end{equation}
where phenomenological constant ${\gamma}$ 
represents the interactions among the 
grain order parameter. We take $\gamma$ to be 1.

%The regular solution parameter incorporates two contributions: one from the chemical effects alone and the other arising from the disparity in atomic sizes between the solute and host atoms (size mismatch). This approach does not consider
%the contribution of ${\Omega}$. Hence, we disregard the non-ideal behaviour and the elastic strain effects in grain boundary segregation.

We define the solute grain boundary interaction potential ${E}$ as 
$-m.\omega.\phi({\eta_1},{\eta_2},..,{\eta_g})$, where the parameter ${m}$ determines the 
strength of the interaction, and $\phi({\eta_1},{\eta_2},..,{\eta_g})$ represents a function derived from the 
grain structure free energy density $g({\eta_1},{\eta_2},..,{\eta_g})$. 
Moreover, $\phi({\eta_1},{\eta_2},..,
{\eta_g})$ is defined as:

%\begin{equation}
       
 %         \phi({\eta_1},{\eta_2},....,{\eta_g})  = \\ 
  %        \frac{A\left[\left|1-\left\{B\left(\left|C\cdot x-0.5\right|\right)\right\}^{N}\right|+1-\left\{B\left(\left|C\cdot x-0.5\right|\right)\right\}^{N}\right]+2}{R}-D.          
   %       \label{eq:g'}
    
%\end{equation}

\begin{multline}
\label{eq:g'}
\phi({\eta_1},{\eta_2},....,{\eta_g})  = \\ \frac{A\left[\left|1-\left\{B\left(\left|C\cdot x-0.5\right|\right)\right\}^{N}\right|+1-\left\{B\left(\left|C\cdot x-0.5\right|\right)\right\}^{N}\right]+2}{R}-D. 
\end{multline}

 The parameters A, B, C, D, N, and R 
 mentioned in Eqn~\eqref{eq:g'} exert 
 control over the behavior of ${\phi
 ({\eta_1},{\eta_2},....,{\eta_g})}$. The correct choice of these parameters provides us with 
 the advantage of controlling the shape of 
 the interaction energy function. Detailed information about all the parameters given in equation~\ref{eq:g'} is provided in Section 1 of supplementary material. Among them, the only variable is the parameter $N$ in our 
simulation, as it directly controls the 
shape of $\phi$ and consequently 
influences the shape of $E$.
 
\begin{figure}[htpb]
\centering 
   \includegraphics[width=1.0\linewidth]{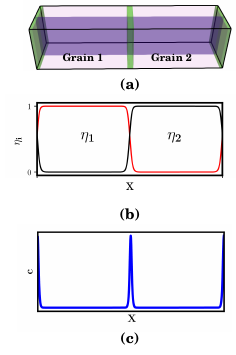}
\caption{(a) Schematic of a bicrystal 
system having flat grain boundary of 
finite thickness. Each grain is 
represneted by light red colored box, 
while the grain boundary is represented as 
the green area.  (b) Variation of grain 
order parameter $\eta_i$. $\eta_1$ 
represents grain 1 and $\eta_2$ represents 
grain 2. (c) Variation of solute 
concentration throughout the system.}
\label{fig:grain_schematic}      % Give a unique label
\end{figure}

In the evolution process, we numerically 
solve the Cahn-Hilliard equation 
(Equation~\ref{eq:seg_5}) for the 
conserved parameter (${c}$)~\cite{CHEQN} and 
the Allen-Cahn equation 
(Equation~\ref{eq:seg_6}) for the non-
conserved parameter (${{\eta_i}}$)
~\cite{ACEQN}, givn below.

\begin{equation}          
          \frac{\partial c}{\partial t} = \nabla\cdot M_c \nabla \Big (\frac{\partial f_{chem}}{\delta c} -m\cdot\omega\cdot \phi-\kappa_c{\nabla}^2c \Big ),   
          \label{eq:seg_5}
\end{equation}

\begin{equation}          
          \frac{\partial \eta_i}{\partial t} = -L \Big (\omega\frac{\partial g}{\partial \eta_i}-m{\cdot}c\cdot\omega\cdot\frac{\partial \phi}{\partial \eta_i}  -\kappa_{\eta}{\nabla}^2\eta_i \Big ).   
          \label{eq:seg_6}
\end{equation}

In these equations, ${M_c}$ represents the 
solute's mobility, ${L}$ is the grain 
boundary relaxation parameter, and ${t}$ 
denotes time. The mobility term ${M_c}$ is 
defined as ${[M \cdot c(1-c)]}$, where 
${M=\frac{D}{RT}}$ and ${D}$ represents 
the interdiffusion coefficient.

The proposed model was implemented using 
the open-source finite element-based 
solver MOOSE Framework
~\cite{lindsay2022moose, 
schwen2023phasefield}. We use periodic boundary condition in the X and Y 
directions. Furthermore, we use Newton's 
method to solve the governing equations in 
MOOSE. Note that, the total composition in 
the entire domain remains conserved 
through the entire span of the simulations.  We employ a computational domain of $4096\Delta X\times128\Delta Y$ in our phase-field simulations. We set the initial solute concentration ($c_0$) to 0.1 for all simulations.
Table~\ref{Tab:param} provides the details about the parameters. The simulation details can be found in Section 2 of the supplementary material. Detailed derviation of the governing equations in MOOSE is provided in Section 3 of supplementary material.

\begin{table}[ht]
\centering
\caption{Parameter details}
  \begin{tabular}{ c c} 
    \toprule
    \multirow{1}{*}{Parameter} 
      & {Value} \\
      \midrule
 $c_o$ & 0.01\\
 $m$ & 8.0\\ 
 $\kappa_{\eta}$ & $1.46\times10^{-11}$ $Jm^{-1}$\\
 $\mu^{o}$ & $1.08\times10^9$ $Jm^{-3}$\\ 
 
 $\mu^{o}_{h}$ & $1.08\times10^9$ $Jm^{-3}$\\ 
 $\omega$ & $1.23\times10^{9}$ $Jm^{-3}$\\
 $M^{o}_{c}$ & $1.70\times10^{-26}$$m^5J^{-1}s^{-1}$\\ 
 $L$ & $2.01\times10^{-5}$ $m^3J^{-1}s^{-1}$\\ 
 $T$ & $655.50$ $K$\\ 
 $\sigma_{gb}$ & $1.0$ $Jm^{-2}$\\  
 $l_{gb}$ & $0.9$ $nm$\\ 
 $\Delta$X & $1.45\times10^{-11}$ $m$\\ 
 $\Delta t$ & $4.57\times10^{-5}$ $s$\\ 
    \bottomrule
  \end{tabular}
  \label{Tab:param}
\end{table}

\begin{figure*}[ht]
    \centering
    \includegraphics[width=1.0\textwidth]{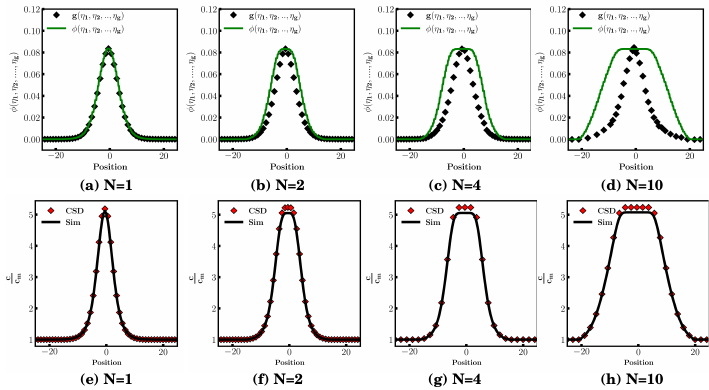}
    \caption{Comparison between $\phi({\eta_1},{\eta_2},..,{\eta_g})$ and $g({\eta_1},{\eta_2},..,{\eta_g})$ for (a) N=1, (b) N=2, (c) N=4 and (d) N=10. Simulated composition profile along with Cahn's solute drag theory (CSD) for stationary GB for (e) N=1, (f) N=2, (g) N=4 and (h) N=10. Here `CSD' refers to Equation~\ref{eq:CSD_static}. }
    \label{fig:static}
\end{figure*}

\section{Results and discussion}
 We start our study by cosidering a 
 bicrystal structure featuring a flat 
 stationary GB (as illustrated in 
 Fig~\ref{fig:grain_schematic}a). Since 
 the grain boundary is flat, there is no 
 curvature effect, resulting in stationary 
 GB. This is intentional, as our objective 
 is to investigate the segregation profile 
 of the grain boundary for various 
 segregation shapes and validate the 
 results with Cahn's theory. As the values 
 of N change, the interaction potential 
 (E) also varies, thereby influencing the 
 segregation shapes at the grain boundary. 
Figures~\ref{fig:static}(a-d) illustrate 
the comparison between ${g}$ and ${\phi}$ 
for various ${N}$ values: 1, 2, 4, and 10. 
Clearly, when ${N=1}$, both ${g}$ and 
${\phi}$ functions are indistinguishable. 
Nevertheless, with increasing ${N}$, the 
${\phi}$ curve widens symmetrically, 
leading to a flattened top section. Figures~\ref{fig:static}(e-h) present the 
corresponding ratio of solute 
concentration ($c$) and matrix solute 
concentration ($c_m$) plot at equilibrium 
and compares with the Cahn Solute Drag 
(CSD) theory~\cite{CAHN1962789}.
As the value of ${N}$ increases, the width 
of ${\phi}$ also broadens (as depicted in 
Figures~\ref{fig:static}(a-d)). 
Consequently, the interaction potential $
E=-m.{\omega}.\phi$ varies with ${N}$, 
affecting the grain boundary's segregation 
profile. To validate our simulated 
segregation profile, we compared
it with Cahn's equation 
(Eq~\ref{eq:CSD_static}) for stationary 
grain boundary as shown :
\begin{equation}
      c=c_m\cdot exp\Big\{\frac{-m.\omega . \phi}{RT}\Big\},
    \label{eq:CSD_static}
\end{equation}
Our simulation results are 
consistent with the segregation theory 
proposed by Cahn~\cite{CAHN1962789} grain 
boundaries. As for stationary GB, the magnitude of flux of solute atoms is identical on both faces of GB, resulting in a symmetrical segregation wrt. the grain GB centre line.

We proceed with simulating various 
segregation shapes in migrating GB and 
examining their impact on the solute drag 
force ($P_{drag}$). Solute drag force 
becomes relevant when GB is in motion. 
However, due to flat 
GB, the GB remains stationary. To overcome this difficulty, we introduce additional energy 
$h(\eta_2)$ (given below)in Eq~\ref{eq:seg_2} based on 
the approach proposed by Heo et 
al.~\cite{HEO20117800}:  

\begin{equation}
      h(\eta_2)=\beta(-2{\eta_2}^3+3{\eta_2}^2).
    \label{eq:seg_11}
\end{equation}

\begin{figure*}[htpb]
\centering 
   \includegraphics[width=0.9\textwidth]{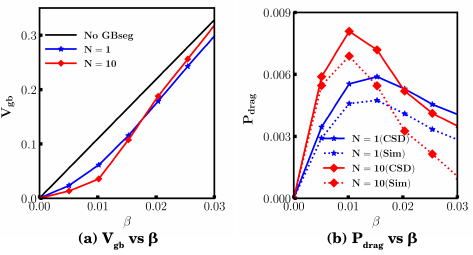}
\caption{(a) Comparision of velocity 
${(V_{gb})}$ vs driving force (${\beta}$) 
for N=1 and 10. (b) Comparison of drag 
force ($P_{drag}$) for N=1 and 10. Here 
`Sim' refers to $P_{drag}$ calculation 
using Equation~\ref{eq:seg_12} and `CSD' 
refers to the calculation using 
Equation~\ref{eq:seg_drag}.}
\label{fig:velocity}      % Give a unique label
\end{figure*}
Here, $\beta$ represents the magnitude of 
the driving force. For $\eta_2=0$, 
$h(\eta_2)=0$, while for $\eta_2=1$, 
$h(\eta_2)=\beta$. A series of simulations are performed for 
a range of driving forces ($\beta$=0.005 
to 0.030).

%Additionally, $\frac{dh}
%{d\eta_2}=0$. This equation introduces 
%additional energy at grain-2 without 
%altering the equilibrium state. 
%By incorporating this feature, it becomes possible to analyze the solute drag effect using different driving forces for grain boundary motion. 

\begin{figure*}[ht]
    \centering
    \includegraphics[width=1.0\textwidth]{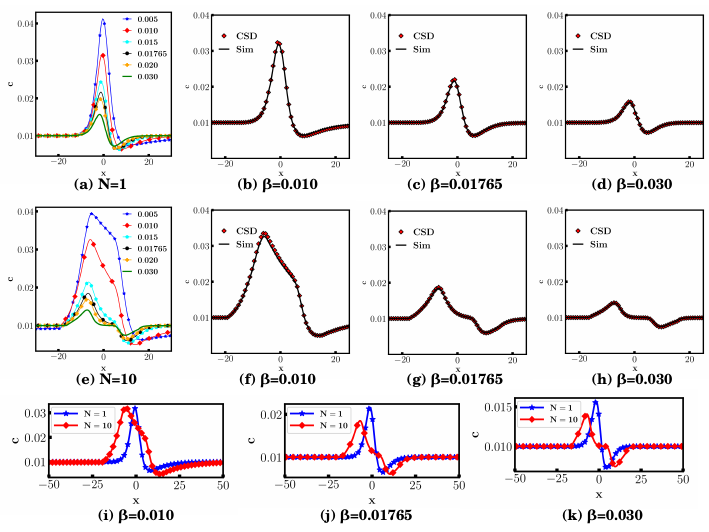}
    \caption{ (a) Evolution of grain boundary segregation with driving force $(\beta)$ for N=1. Comparison of segregation profile with Cahn's solute drag theory (CSD) for (b) $\beta=0.010$, (c) $\beta=0.01765$ and (d) $\beta=0.030$, respectively. (e) Evolution grain boundary segregation with driving force $(\beta)$ for N=10. Comparision of segregation with CSD for (f) $\beta=0.010$, (g) $\beta=0.01765$ and (h) $\beta=0.030$, respectively. Comparison of segregation profile between N=1 and N=10 for (i) $\beta=0.010$, (j) $\beta=0.01765$ and (k) $\beta=0.030$, respectively. Here, `CSD' refers to calculation using Equation~\ref{eq:CSD_moving}.   }
    \label{fig:moving}
\end{figure*}

For migrating GB, we conduct simulations 
for both without segregation and with 
segregation having different shapes 
(N=1,2,4 and 10). Figure~\ref{fig:velocity}a illustrates the 
relationship between the grain boundary 
velocity ($V_{gb}$) and the driving force 
($\beta$). The black solid line represents 
the velocity profile without any grain 
boundary segregation. The relationship 
between the grain boundary velocity and 
the driving force can be expressed as 
follows:

\begin{equation}
      V_{gb}=M_o(\beta-P_{drag}).
    \label{eq:seg_12}
\end{equation}

The equation includes $M_o$, representing 
the intrinsic mobility of GB. With no 
solute segregation ($P_{drag}=0$), the 
simplified equation is 
$V_{gb}=M_o{\beta}$. In the absence of 
segregation, GB velocity shows a linear 
relationship with the driving force, 
yielding $M_0 = 10.9$ in dimensionless 
units from the slope of the plot.

For grain boundaries (GB) with solute 
segregation, two distinct velocity regimes 
emerge: an initial low-velocity regime 
followed by a later high-velocity regime 
for all segregation shapes (shown by blue 
(N=1) and red (N=10) lines in 
Figure~\ref{fig:velocity}a). The velocity 
vs. driving force plot for all segregation 
patterns is given in Section 4 of supplementary material.  
During the low-velocity regime, N=1 
exhibits higher velocity than N=10. 
However, in the high-velocity regime, the 
velocity surpasses a certain point. The velocity crossover between 
N=1 and N=10 occurs at $\beta=0.01765$. 
After the crossover, N=1 exhibits lower velocity 
than N=10. 
Figure~\ref{fig:velocity}(b) shows the 
comparison between drag force calculated 
using Eq~\ref{eq:seg_12} and 
Equation~\ref{eq:seg_drag}~\cite{HEO20117800} 
for N=1 and N=10. At low velocity, N=10 
experiences higher drag force, while at 
high velocity, it encounters lower drag 
force compared to N=1. The crossover of 
drag force found to be around $\beta=0.01765$.

\begin{equation}
                P_{drag} = RTV_{gb}\int_{-\infty}^{+\infty}\frac{(c-c_m)}{D(1-c)}dx. 
    \label{eq:seg_drag}
\end{equation}

Figure~\ref{fig:moving}(a) displays the 
composition plot for N=1 at various 
driving forces. A dip in solute 
concentration is observed in front of the 
moving grain boundary, consistent with 
experimental 
findings~\cite{MAVRIKAKIS2020100541, 
D0NR07180C}. A similar concentration dip 
is evident for N=10 (shown in 
Figure~\ref{fig:moving}(e)). As the 
driving force increases, the grain 
boundary moves faster, making segregation 
challenging for solute atoms. Thus, peak 
segregation decreases with higher driving 
forces in both cases. The intensified 
driving force accelerates the grain 
boundary's motion, hindering solute atoms 
from segregating at GB, as 
illustrated in Figure~\ref{fig:moving}a 
and Figure~\ref{fig:moving}e.
We compared the segregation profile and 
CSD~\cite{CAHN1962789} theory for migrating 
grain boundary with constant diffusivity, 
as described by: 

\begin{equation}
      D\frac{\delta ^2c}{\delta x^2}+\Big\{\frac{D}{RT}\frac{\delta E}{\delta x} +V_{gb} \Big \}\frac{\delta c}{\delta x} + \frac{c}{RT}\Big\{D\frac{\delta ^2E}{\delta x^2} \Big\} = 0
    \label{eq:CSD_moving}
\end{equation}

Here, the interaction potential term is 
$-m.{\omega}.\phi$. Figure~\ref{fig:moving}(b-
d) depicts the comparison of solute 
concentration between our simulation and 
the CSD theory 
(Equation~\ref{eq:CSD_moving}) for N=1 
under different driving forces of 0.010, 
0.01765, and 0.030. Figure
s~\ref{fig:moving}(f-h) depict the 
comparison of solute concentration between 
our simulation and the CSD theory 
(Equation~\ref{eq:CSD_moving}) for N=10 
under different driving forces of 0.010, 
0.01765, and 0.030. Our simulation results agrees well with the predictions from
the CSD theory for both N=1 and N=10.

For values of $N>1$, indicate the peak 
segregation occurs across a finite region 
(shown in Figures~\ref{fig:static}(f-h)) 
and the flat region of solute segregations 
observed for a static GB becomes 
inclined towards the direction of GB 
migration. \textcolor{black}{In the case of a stationary 
GB, solute atoms from the adjacent matrix 
on both sides of the GB migrate towards it at a same rate, leading to their accumulation at GB. This atomic migration continues until GB segregation reaches a state of equilibrium with matrix.} Consequently, the flux on 
both sides of the grain boundary is equal 
($|J_f^-| = |J_f^+|$) where $J_f$ is Fickian 
flux.

For a moving GB, the amount of segregation diminishes with the increase in the driving force for GB migration, which lowers the peak segregation. The $x-$component of the total atomic flux with reference to a moving boundary is expressed as $J = J_f - V_{gb}\cdot c$ where the coordinate system is attached to the moving GB and $V_{gb}$ is a positive quantity.

The magnitude of the x-component flux is substantially higher when emanating from the front face of GB, whereas it is comparatively lower from the rear face of GB. In this context, the rear face registers a positive value for $J_f$ due to atom movement directed towards the grain boundary's center from the rear face (in the positive $x-$direction). Conversely, the front face presents a negative $J_f$ as atoms migrate in the opposing direction.

%Here So the magnitude of the $x-$component of flux from the front face is high while that from the back face is low. Here, $J_f$ is positive for the back face as the movement of atoms is towards the center of the gb from the back face (towards positive $x-$direction), while $J_f$ is negative for the front face as the atoms move in the opposite direction. 

Thus, solute atoms undergo longer transportation distances towards the left from the front face, while their motion from the back face is confined to shorter distances towards the right. This results in the accumulation of solute atoms alongside the right side of the back face. Consequently, more segregation of solute atoms adjacent to the back face is evident compared to the front face, leading to a solute concentration gradient within GB. However, an escalation in the grain boundary migration velocity occurs with an increase in the driving force. This, in turn, reduces the total time required for solute atoms to segregate, leading to a decrease in the overall extent of segregation.

It is observed that the amount of solute segregation decreases 
with increasing driving force, as shown in 
Figure~\ref{fig:moving}(e). As the driving 
force increases, the flux at the front 
of GB decreases while the flux behind the GB increases. Consequently, the 
difference in solute concentration between 
the two faces becomes more pronounced. 
With further increase in velocity, the 
flux on the front face approaches zero, 
resulting in negligible segregation on the 
front face, as shown in the Fig
ure~\ref{fig:moving}e for $\beta=0.030$.

%As the velocity of the grain boundary increases, the flux through the grain boundary also increases, which can be represented by the equation J = $V_{gb} \cdot A_{gb}$, where J denotes flux wrt. static frame of reference, $V_{gb}$ represents grain boundary velocity, and $A_{gb}$ represents the grain boundary area. In our case, with constant solute diffusivity, an increase in flux leads to an increase in the concentration gradient. This can be understood by the equation J = $-D \cdot \frac{\partial C}{\partial X}$, where D denotes diffusivity, and $\frac{\partial C}{\partial X}$ represents the concentration gradient. Consequently, as the driving force increases, the concentration gradient within the grain boundary intensifies. 

Figures~\ref{fig:moving}(i-j) illustrate 
the comparison of solute concentration 
profiles for different driving forces of 
0.010, 0.01765, and 0.030. Figure~\ref{fig:moving}i focuses explicitly on the 
driving force of 0.010, representing the 
pre-crossover stage. In this case, it is 
evident that the total solute level is 
higher for N=10 compared to N=1. For 
$\beta=0.01765$ (shown in 
Figure~\ref{fig:moving}j), the total 
solute amount is approximately the same 
for both N=1 and N=10. However, for 
$\beta=0.030$ (shown in 
Figure~\ref{fig:moving}k), the total 
solute amount is lower for N=10 than N=1. 
The fact that the drag force for a moving 
grain boundary is a function of $\int_{-
\infty}^{+\infty}\frac{(c-c_m)}{D(1-c)}$ 
justifies the occurrence of the crossover 
between N=1 and N=10 under the driving 
force.

\section{Conclusions}
\textcolor{black}{We present a novel phase field model for simulating GB segregation, highlighting varied segregation patterns through modified solute-GB interactions. We alter GB segregation profiles by adjusting the parameter of $N$.
For $N>1$, peak segregation spans over a finite region. An equal flux of solute atoms from both faces of GB leads to symmetrical solute segregation in stationary GB.
However, in the case of a moving GB, the atomic flux is higher at the front face, resulting in more segregation near the back face, establishing a solute concentration gradient within the GB. 
Increasing driving force escalates GB velocity, reducing solute atoms accumulation time, hence lowering segregation.
Additionally, we observe that in the low-velocity regime, $N=1$ exhibits faster GB velocity than $N>1$, while the reverse holds in the high-velocity regime.
Our phase-field simulations align with Cahn's classical theory on solute drag for both stationary and moving GB.  }

\section*{Acknowledgments}

One of the authors R.M. acknowledge 
financial support from SERB core research 
grant ( CRG/2019/006961 ). The authors 
acknowledge National Supercomputing 
Mission (NSM) for providing computing 
resources of ‘PARAM Sanganak’ at IIT 
Kanpur, which is implemented by C-DAC and 
supported by the Ministry of Electronics 
and Information Technology (MeitY) and 
Department of Science and Technology 
(DST), Government of India. 

\section*{Data Availability Statement}
The data that support the findings of this 
study are available from the corresponding 
author upon reasonable request.

\section*{AUTHOR CONTRIBUTIONS}
\textbf{Sandip Guin:} Conceptualization, Visualization, Methodology, Software, Investigation, Formal analysis, Validation, Data curation, Writing-Original Draft. \textbf{Miral Verma:} Methodology, Formal analysis, Investigation, Writing-Original Draft. \textbf{Soumya Bandyopadhyay :} Methodology, Formal analysis, Investigation, Software, Writing-Original Draft. \textbf{Yu-Chieh Lo:} Supervision, Project administration, Resources, Writing - review $\&$ editing. \textbf{Rajdip Mukherjee:} Supervision, Project administration, Resources, Writing - review $\&$ editing, Funding acquisition.

\bibliographystyle{unsrt}
\bibliography{mybibfile}

\end{document}

\begin{frontmatter}
\title{\textbf{Supplementary Material}\\ Solute Segregation in a Moving Grain Boundary: A Novel Phase-Field Approach
}
\author[a,b,d]{Sandip Guin}
\author[c]{Miral Verma}
\author[a]{Soumya Bandyopadhyay}
\author[b,d]{Yu-Chieh Lo*\corref{cor}}
\ead{yclo@nycu.edu.tw}
\author[a]{Rajdip Mukherjee*\corref{cor}}
\ead{rajdipm@iitk.ac.in}
\address[a]{Department of Materials Science and Engineering, Indian Institute of
Technology, Kanpur, Kanpur-208016, UP, India}
\address[b]{International College of Semiconductor Technology, National Yang Ming Chiao Tung University, Hsinchu 300, Taiwan} 
\address[c]{Department of Materials Engineering, KU Leuven, Kasteelpark Arenberg 44, Leuven 3001, Belgium}
\address[d]{Department of Materials Science and Engineering, National Yang Ming Chiao Tung University, Hsinchu 300, Taiwan}

\end{frontmatter}

\linenumbers

\section{Interaction potential}
\label{phidetails}

In our phase-field simulation, we use solute GB interaction potential as $-m.\omega .\phi$, where $\phi$ is given by:

\begin{equation}
          \begin{aligned}
          \phi({\eta_1},{\eta_2},....,{\eta_g})  &= 
          \frac{A\left[\left|1-\left\{B\left(\left|C\cdot x-0.5\right|\right)\right\}^{N}\right|+1-\left\{B\left(\left|C\cdot x-0.5\right|\right)\right\}^{N}\right]+2}{R}-D.
          \end{aligned}
          \label{eq:phi}
\end{equation}

Here the parameters A, B, C, D, N, R are constant. By changing the values of this parameter we can change the shape of the interaction potential, hence the GB segregation shape.

\begin{figure}[htpb]
\centering 
   \includegraphics[width=1.0\textwidth]{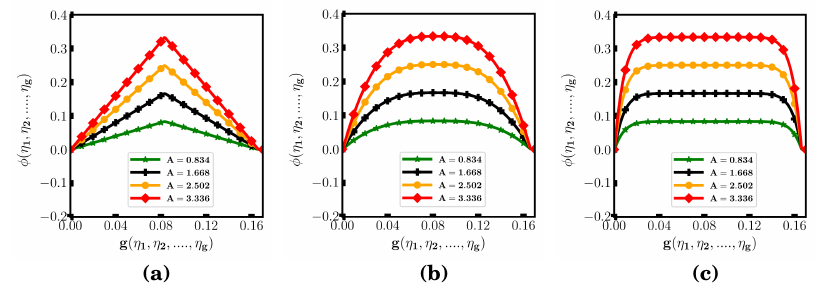}
\caption{$\phi(\eta_i)$ vs $g(\eta_i)$ plot for different A for (a)N=1, (b)N=3, and (c)N=10.}
\label{fig:a_varr}      % Give a unique label
\end{figure}

With an increase in the values of A, the peak value of the function $\phi$ wrt. $g$ increases while, the lower value of  $\phi$ for same $g$ remains the same. Figures~\ref{fig:a_varr}(a-c) depict the $\phi$ vs. $g$ plot, demonstrating how it varies with different A values at N=1 (Figure~\ref{fig:a_varr}a), N=3 (Figure~\ref{fig:a_varr}b), and N=10 (Figure~\ref{fig:a_varr}c). Moreover, as N increases, the upper section of the curve tends to flatten out.

\begin{figure}[htpb]
\centering 
   \includegraphics[width=1.0\textwidth]{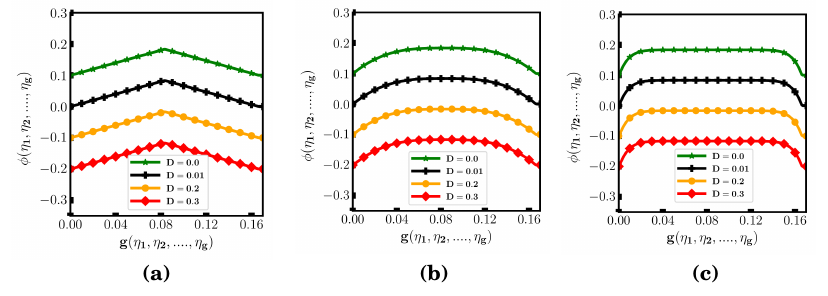}
\caption{$\phi(\eta_i)$ vs $g(\eta_i)$ plot for different D for (a)N=1, (b)N=3, and (c)N=10.}
\label{fig:D_varr}      % Give a unique label
\end{figure}

As D values increase, the entire $\phi$ vs. $g$ plot shifts downwards. Figures~\ref{fig:D_varr}(a-c) illustrate the variation of the $\phi$ vs. $g$ plot with different D values for N=1 (Figure~\ref{fig:D_varr}a), N=3 (Figure~\ref{fig:D_varr}b), and N=10 (Figure~\ref{fig:D_varr}c). Additionally, as N increases, the upper section of the curve tends to flatten out, similar to the previous case.

\begin{figure}[htpb]
\centering 
   \includegraphics[width=1.0\textwidth]{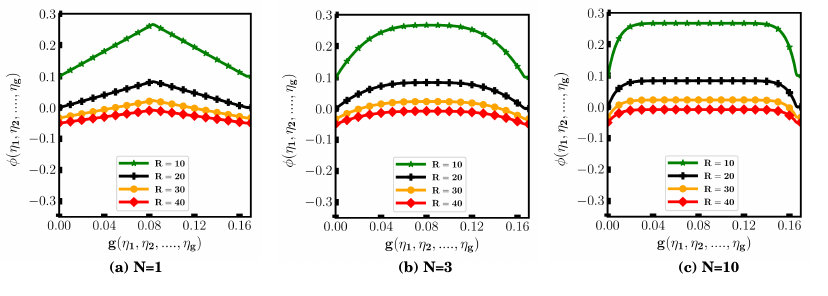}
\caption{$\phi(\eta_i)$ vs $g(\eta_i)$ plot for different R for (a)N=1, (b)N=3, and (c)N=10.}
\label{fig:R_varr}      % Give a unique label
\end{figure}

As R values increase, the entire $\phi$ vs. $g$ plot shifts downwards. Figures~\ref{fig:R_varr}(a-c) illustrate the variation of the $\phi$ vs. $g$ plot with different R values for N=1 (Figure~\ref{fig:R_varr}a), N=3 (Figure~\ref{fig:R_varr}b), and N=10 (Figure~\ref{fig:R_varr}c). Additionally, as N increases, the upper portion of the curve tends to flatten out, similar to the previous cases.

\begin{figure}[htpb]
\centering 
   \includegraphics[width=1.0\textwidth]{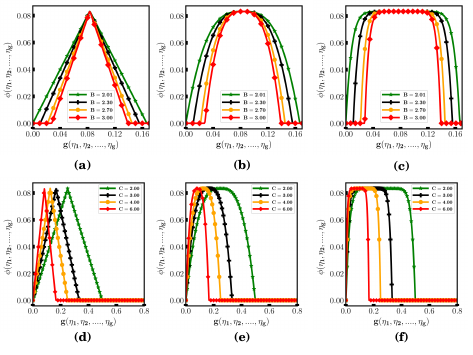}
\caption{$\phi(\eta_i)$ vs $g(\eta_i)$ plot for different B for (a)N=1, (b)N=3, and (c)N=10. $\phi(\eta_i)$ vs $g(\eta_i)$ plot for different C for (a)N=1, (b)N=3, and (c)N=10. }
\label{fig:B_c_var}      % Give a unique label
\end{figure}

Figures~\ref{fig:B_c_var} display the variation of $\phi$ for different values of ${B}$ and ${C}$ with respect to $g$. In Figure~\ref{fig:B_c_var}(a-c), the curves represent the relationship between $\phi$ and $g$ for varying B, C = 6.0, D = 0.1, R = 20.0, and (a) N = 1, (b) N = 3, (c) N = 10. It is evident that as the value of ${B}$ increases, the width of the curve symmetrically expands keeping the centre value at the same position. An important observation is that the maximum value of $\phi$ is located at the center position. However, as the value of C increases, the $\phi$ curve shifts towards the left side, maintaining a constant maximum value (shown in Figures~\ref{fig:B_c_var}(d-f)). Moreover, in both cases, an increase in N causes the flattening of the upper part of the curve. 

Heo et al. employed E as  $-m.\omega.{g}$~\cite{HEO20117800} where, the value of ${g}$ is zero inside the matrix, and gradually increases within GB, and reaches its maximum at the center of the GB. Consequently, the interaction energy in their work follows a similar pattern of variation. 

\begin{figure}[htpb]
\centering 
   \includegraphics[width=0.7\textwidth]{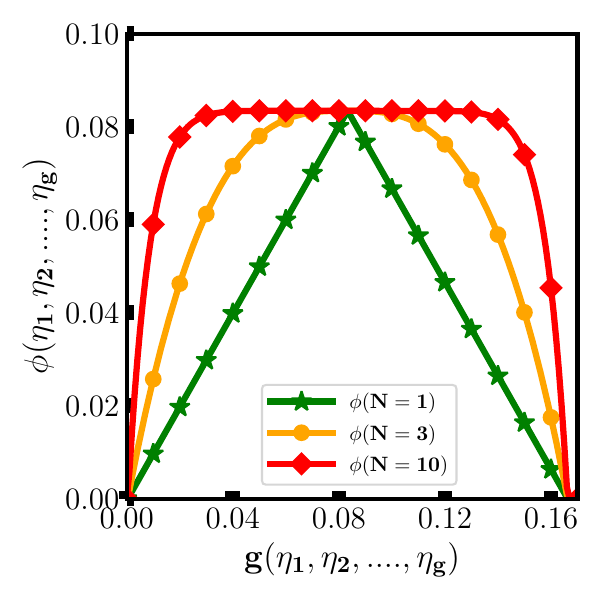}
\caption{$\phi(\eta_i)$ vs $g(\eta_i)$ plot for N=1, 3 and 10.}
\label{fig:N_varr}      % Give a unique label
\end{figure}

In our work, however, we use E as
$-m.\omega.{\phi}$ instead of 
$-m.\omega.{g}$, which provides us with 
greater control over the shape of the 
interaction energy. Thus, it is crucial to 
appropriately select all the parameters in 
such a way that ${\phi}=0$ inside the 
matrix, at the center of the GB, both 
$\phi$ and $g$ acquire the same value, and 
for N=1, both functions are identical. For 
this reason, we choose B=2.01, C=6.0, 
R=20.0, ${A=\frac{R}{2}.g({\eta_1},
{\eta_2},....,{\eta_g})|_{x= \text{GB 
center}}}$ and $D=\frac{2}{R}$. Furthermore, keeping all the other parameters fixed, we only vary N in our simulation to control the shape of interaction potential function; the shape of GB solute segregation. We observe, as  value of N increases,  the upper portion
of the curve tends to flatten. 
Figure~\ref{fig:N_varr} shows the $\phi$ 
vs $g$ plot for different values of N.

%The parameter ${A}$ is associated with the grain structure energy value at the grain boundary's centre point. Specifically, ${A}$ is defined as ${[p.g({\eta_1},{\eta_2},....,{\eta_g})|_{x= \text{GB center}}]}$. We consider a bicrystal system (depicted in Figure~\ref{fig:grain_schematic}a) consisting of a single flat grain boundary. Each grain is represented by ${\eta_i}$ (as shown in Fig~\ref{fig:grain_schematic}b). We define intersection point between 1D ${\eta_1}$ and ${\eta_2}$ as the middle of GB. Fig~\ref{fig:grain_schematic}c shows the variation of solute concentration throughout the system.  The parameter ${D}$ is assigned a value of 0.1, and ${R}$ is set to 20. These values are chosen such that the value of ${g'({\eta_1},{\eta_2},....,{\eta_g})}$ within the grain is zero. The parameter ${N}$ is related to the shape of the interaction potential between the solute and grain boundary. Further information about the parameter N is provided in the supplementary document.

\newpage
\section{Simulation details}
\label{simd}
For our phase-field simulations, we use a computational domain of $4096\Delta X\times128\Delta Y$ with periodic boundary conditions in both the X and Y directions. The initial concentration of the solute ($c_0$) for all 
simulations was set to 0.1. We considerd, the grain boundary energy ($\sigma_{gb}$) to be 1.0 $Jm^{-2}$, and the measured grain boundary width ($l_{gb}$) as 0.9 $nm$. The molar volume ($V_m$) of the material was 11.0 $cm^{3}mol^{-1}$. Please 
refer to Table~\ref{parameters_table} for further information regarding the simulation parameters. We use all the parameters in our simulations in their corresponding non-dimensional form. To perform non-dimensionalization,  in our 
case, the characteristic energy ($E_c$) is $1.08
\times10^9$ $Jm^{-3}$, and the characteristic length ($l_c$) is $2.90\times10^{-11}$ m. For a comprehensive 
understanding of the dimensional conversion, we refer to the article by Heo et al~\cite{HEO20117800}.  

\begin{table*}[htbp]
\begin{center}
\caption{Parameters used in phase-field simulations}
\begin{tabular}{ |c|c|} 
%\caption{Parameters Details}
 \hline
 Parameter & Values \\ 
 \hline

 $c_o$ & 0.01\\
 $m$ & 8.0\\ 
 $\kappa_{\eta}$ & $1.46\times10^{-11}$ $Jm^{-1}$\\ 
 $\mu^{o}$ & $1.08
\times10^9$ $Jm^{-3}$\\ 

 $\mu^{o}_{h}$ & $1.08
\times10^9$ $Jm^{-3}$\\ 

 $\omega$ & $1.23\times10^{9}$ $Jm^{-3}$\\
 
 $M^{o}_{c}$ & $1.70\times10^{-26}$ $m^5J^{-1}s^{-1}$\\ 
 
 $L$ & $2.01
\times10^{-5}$ $m^3J^{-1}s^{-1}$\\ 

 $T$ & $655.50
$ $K$\\ 

 $\sigma_{gb}$ & $1.0$ $Jm^{-2}$\\ 
 
 $l_{gb}$ & $0.9$ $nm$\\
 
 $\Delta$X & $1.45\times10^{-11}
$ $m$\\ 
 $\Delta t$ & $4.57\times10^{-5}$ $s$\\ 
 \hline
\end{tabular}
\end{center}
\label{parameters_table}
\end{table*}

%\begin{figure}[htpb]
%\centering 
%   \includegraphics[width=0.9\linewidth]{images/velocity.pdf}
%\caption{Velocity vs driving force plot for various N. }
%\label{fig:2}      % Give a unique label
%\end{figure}

\newpage
\section{Finite element discretization of the governing equations in MOOSE Framework}
\label{fem}
The finite element method (FEM) has been widely used to solve complex partial differential equations (PDE)
with ease for many years by the scientific community due to simplicity in the discretization of the equations
into the weak form and efficiently handle the exact boundary conditions.
Here we perform a FEM-based method to solve the 
governing phase-field equations (Eqs.~\eqref{eq:seg_5} \&~\eqref{eq:seg_6}) with periodic boundary conditions, 

\begin{equation}          
          \frac{\partial c}{\partial t} = \nabla\cdot M_c \nabla \Big (\frac{\partial f_{chem}}{\delta c} -m\cdot\omega\cdot \phi-\kappa_c{\nabla}^2c \Big ),   
          \label{eq:seg_5}
\end{equation}

\begin{equation}          
          \frac{\partial \eta_i}{\partial t} = -L \Big (\omega\frac{\partial g}{\partial \eta_i}-m{\cdot}c\cdot\omega\cdot\frac{\partial \phi}{\partial \eta_i}  -\kappa_{\eta}{\nabla}^2\eta_i \Big ).   
          \label{eq:seg_6}
\end{equation}

Solving the Cahn-Hilliard equation is a bit difficult, 
due to the presence of gradient-free energy which is a fourth-order differential equation. But, it can be solved in two ways. 
First to prepare for the FEM discretization of Eqn.~\eqref{eq:seg_5}, 
we construct a weak form in a similar manner to that used by Stogner
et al.~\cite{stogner2008approximation}. The weighted integral residual projection of Eqn.~\eqref{eq:seg_5} 
is constructed using a test function $\phi_m$ and integrating
the second-order terms by parts once and the fourth order term
by parts twice. Thus after discretization Eqn.~\eqref{eq:seg_5} yields:
\begin{equation}
    \begin{split}
        \left(\frac{\partial c(\mathbf{r},t)}{\partial t}, \phi_{m}\right) &= -\left(\kappa_{c} \nabla^{2}c(\mathbf{r},t), \nabla\cdot(M_{c}(c(\mathbf{r},t), \theta)\nabla\phi_{m})\right)\\
                                                           &-\left(M_{c}(c(\mathbf{r},t), \theta) \nabla \left(\frac{\partial f(c(\mathbf{r},t)}{\partial c(\mathbf{r},t)}\right), \nabla \phi_{m}\right)\\
                                                           & +\langle M_{c}(c(\mathbf{r},t), \theta) \nabla(\kappa_{c}(\nabla^2 c(\mathbf{r},t))\cdot \hat{n}), \phi_m \rangle\\
                                                           &- \left\langle M_{c}(c(\mathbf{r},t), \theta)\nabla \left(\frac{\partial f(c(\mathbf{r},t)}{\partial c(\mathbf{r},t)}\right)\cdot \hat{n}, \phi_m \right \rangle\\
                                                           &+\langle \kappa_{c}\nabla^2 c(\mathbf{r},t), M_{c}(c(\mathbf{r},t), \theta)\nabla\phi_m\cdot \hat{n} \rangle,
    \end{split}
    \label{eqn:16}
\end{equation}
where $(*,*)$ operator represents a volume integral with an inner product and $\langle *,* \rangle$ operator denotes the
surface integral with an inner product. 

  Another way to solve Eqn.~\eqref{eq:seg_5} is to split
the fourth order equation into two second order equations, such that two variables are solved, the concentration $c(\mathbf{r},t)$  
and the chemical potential $\mu$.
In this case, the two residual equations are: 
\begin{subequations}
\begin{equation}
\begin{split}
    R_{\mu} &= \left(\frac{\partial c(\mathbf{r},t)}{\partial t}, \phi_m\right) + \left( M_{c}(c(\mathbf{r},t)\nabla\mu, \mathbf{\nabla} \phi_m \right) \\
          &- \langle  M_{c}(c(\mathbf{r},t)\nabla\mu \cdot\hat{n},\phi_m \rangle. 
   \label{eqn:17a}
   \end{split}
\end{equation}
\begin{equation}
  \begin{split}
     R_{c} &= (\nabla c(\mathbf{r},t),\nabla(\kappa_{c} \phi_m)) - \langle \nabla c(\mathbf{r},t)\cdot\hat{n}, \kappa_{c}\phi_m\rangle \\
               &-\left(\left(\frac{\partial f(c(\mathbf{r},t)}{\partial c(\mathbf{r},t)} - \mu\right),\phi_m\right). 
\end{split}
\label{eqn:17b}
\end{equation}
\label{eqn:17}
\end{subequations}
In this work, we mainly adopt the second or the split formalism of Eqn.~\eqref{eqn:17a} $\&$ Eqn.~\eqref{eqn:17b} 
as this change improves the solve convergence without any impact on the solution.
In a similar manner, we can also construct the weak form and the corresponding residual for Eq.~\eqref{eq:seg_6}~\cite{lindsay2022moose,tonks2012object}. For further details related to the weak form please refer to the MOOSE official website https://mooseframework.inl.gov/.

\section{Velocity profiles}

\begin{figure}[htpb]
\centering 
   \includegraphics[width=0.5\textwidth]{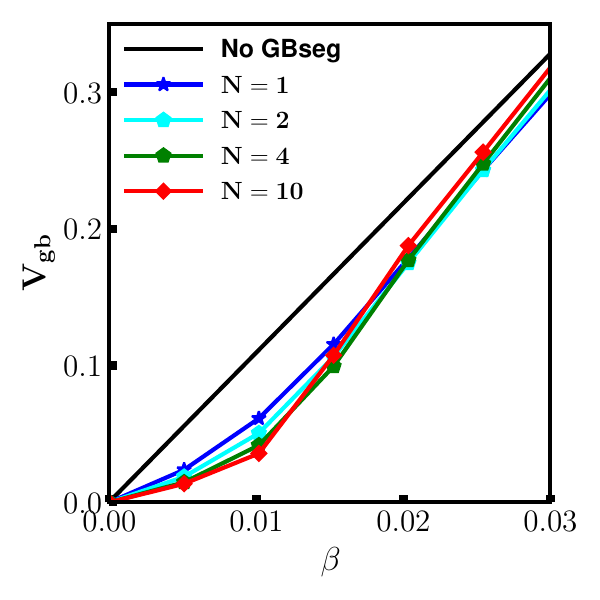}
\caption{Velocity ($V_{gb}$) vs. driving force ($\beta$) for N = 1, 2, 4, and 10}
\label{fig:supl_vel}      % Give a unique label
\end{figure}

%\appendix

\newpage
%\section*{References}
%\bibliographystyle{model1a-num-names}
\bibliographystyle{unsrt}
\bibliography{mybibfile}